\documentclass[nofootinbib,aps,twocolumn]{revtex4}
\usepackage{graphicx}
\usepackage{mathrsfs}
\usepackage{tipa}
\usepackage{bm}
\usepackage{amsmath}
\usepackage{amssymb}
\usepackage{amsthm}
\usepackage{hyperref}

\usepackage{amsfonts}
\usepackage{mathtools}
\usepackage{latexsym}
\usepackage[greek,english]{babel}
\usepackage{booktabs}
\usepackage[iso-8859-7]{inputenc}
\usepackage{color}

\begin{document}

\title{Primordial black holes in a dimensionally oxidizing Universe}

\author{Konstantinos F. Dialektopoulos}
\email{kdialekt@gmail.com}
\affiliation{Center for Gravitation and Cosmology, College of Physical Science and\\ Technology, Yangzhou University, Yangzhou 225009, China}

\author{Piero Nicolini}
\email{nicolini@fias.uni-frankfurt.de}
\affiliation{Frankfurt Institute for Advanced Studies (FIAS), Ruth-Moufang-Str. 1, 60438 Frankfurt am Main, Germany}
\affiliation{Institut f\" ur Theoretische Physik, Goethe Universit\" at Frankfurt, Max-von-Laue-Str.1, 60438 Frankfurt am Main, Germany}

\author{Athanasios G. Tzikas}
\email{tzikas@fias.uni-frankfurt.de}
\affiliation{Frankfurt Institute for Advanced Studies (FIAS), Ruth-Moufang-Str. 1, 60438 Frankfurt am Main, Germany}
\affiliation{Institut f\" ur Theoretische Physik, Goethe Universit\" at Frankfurt, Max-von-Laue-Str.1, 60438 Frankfurt am Main, Germany}

\date{\today}

\begin{abstract}
The spontaneous creation of primordial black holes in a violently expanding Universe is a well studied phenomenon. Based on quantum gravity arguments, it has been conjectured that the early Universe might have undergone a lower dimensional phase before relaxing to the current ($3+1$) dimensional state. 
In this article we combine the above phenomena: we calculate the pair creation rates of black holes nucleated in an expanding Universe, by assuming a dimensional evolution, we term ``oxidation'', from ($1+1$) to ($2+1$) and finally to ($3+1$) dimensions. Our investigation is based on the no boundary proposal that allows for the construction  of the required gravitational instantons. If, on the one hand, the existence of a dilaton non-minimally coupled to the metric is necessary for black holes to exist in the ($1+1$) phase, it becomes, on the other hand, trivial in  ($2+1$) dimensions. Nevertheless, the dilaton might survive the oxidation and be incorporated in a modified theory of gravity in ($3+1$) dimensions: by assuming that our Universe, in its current state, originates from a lower-dimensional oxidation, one might be led to consider  the pair creation rate in a sub-class of the Horndeski action. Our findings for this case show  that, for specific values of the Galileon coupling to the metric, the rate can be unsuppressed. This would imply the possibility of compelling parameter bounds for non-Einstein theories of gravity by using the spontaneous black hole creation. 
\end{abstract}

\maketitle

\section{Introduction}
\label{sec:intro}

Gravity is by far the less understood of the fundamental interactions governing our Universe. This is not only due to its nonlinear nature and the presence of curvature singularities, but also to the difficulty of determining basic parameters. For instance, the value of the coupling constant $G_\mathrm{N}$ is known with an accuracy that has barely improved since the time of Cavendish \cite{BlN14,AHH+07,AGH+09}. Furthermore, gravity resists a direct quantization due to its non-renormalizable character and is nested in most of the ongoing issues that prevent a satisfactory understanding of the Universe. Such issues are, for instance, the transplanckian problem, the hierarchy problem, the nature of dark components, the thermodynamic description of event horizons and their informational content \cite{tGR+18}.

As a viable solution to address some of the above issues, it has been postulated that the actual number of spacetime dimensions might differ from four. This led to the formulation of higher dimensional scenarios, most notably the large extradimension scenario \cite{Ant90,AAD+98,ADD98,ADD99}, the warp geometry models \cite{Gog98a,Gog98b,Gog99,RaS99a,RaS99b}, and the universal extradimensions \cite{ACD01}. Alternatively, a radically opposite idea has been put forward by 't Hooft \cite{tHo93}. In the extreme high energy regime, the spacetime might undergo a dimensional reduction. This feature would allow for an improvement or even a solution of the nonrenormalizability problem of gravity. 
For these reasons, there have been an array of investigations aiming to show the existence of the `dimensional flow' of a spacetime manifold \cite{AJL05,Car09,Ben09,MoN10,MuS11,NiS11,Cal12,MuN13,AAG+13,ASZ14,Car15,Car17,Anchordoqui:2010er,Stojkovic:2014lha}. Rather than a smooth differential manifold, the spacetime at the shortest scale would behave like a fractal due to its fluctuating quantum character. Fractals have a continuous dimension that differs from the topological dimension of the ambient space.  In the case of gravity, the dimensional flow refers to the dependence  of the fractal dimension on the energy at which the spacetime is probed. 
Although the spacetime topological dimension remains unaltered, i.e. four, some phenomena might behave as dimensionally reduced systems, being subject to a loss of local resolution at short scales.

Along this line of reasoning, the repercussions of the dimensional flow have recently been considered by studying the quantum decay of de Sitter space within a two dimensional dilaton gravity formulation \cite{TNM++18}. Although classically stable, de Sitter space is quantum mechanically unstable, resulting in the nucleation of black hole pairs
\footnote{By  pair one means a topological black hole pair.  The Schwarzschild-de Sitter solution is a special case of the known C-metric \cite{KiW70,GKP06}, which represents a pair of causally separated black holes accelerating in opposite directions under the action of forces that are represented by conical singularities (see also \cite{DiL03,KrP03}). Only one  black hole exists inside the observable Universe while the other one is outside the cosmological horizon.}. 
At this point, we recall that primordial black holes (PBHs) are theoretical objects with masses ranging from Planck relics $(\sim 10^{-8}\text{kg})$ up to thousands of  solar masses. PBHs are unlikely to form from the gravitational collapse of a star today, since low-mass black holes can form only if matter is compressed to enormously high densities by very large external pressures \cite{CaH74} and \cite{Nojiri:2020tph,Elizalde:1999dw,Nojiri:1998ph}. Such conditions of high temperatures and pressures can be found in the early stages of a violent Universe and it is believed that PBHs may have been produced plentiful back then. 

According to the instanton formalism for Euclidean quantum gravity proposed by Mann \& Ross \cite{MaR95}, Bousso \& Hawking \cite{BoH95,BoH96},  black holes are copiously produced in ($3+1$) dimensions provided the cosmological constant has Planckian values \cite{MaN11}. Therefore, these nucleated black holes can  be considered ``primordial", relative to those stemming from alternative formation mechanisms \cite{Car05,Khl10}.
As an additional feature, the de Sitter space quantum instability may lead to the resolution of one of aforementioned ongoing issues in physics, since stable relics of primordially produced black holes have been considered as a cold dark matter component \cite{Carr19,Dym20}. This possibility is of primary interest as long as attempts of direct detection of dark matter candidates at the LHC have not found experimental corroboration \cite{Schumann19}.

 There is, however, a potential problem within such a scenario. Black holes produced prior inflation would be washed out with no significant effects on the current Universe \cite{deFS20}. Conversely,  the production of primordial black holes in a dimensionally reduced Universe can occur without requiring Planckian values for the cosmological constant \cite{TNM++18}. This is a clear advantage that descends from the fact that in ($1+1$) dimensions the gravitational coupling is dimensionless and no actual gravitational scale is associated to it. Interestingly such ($1+1$) dimensional black holes have a peculiar thermodynamics. They radiate with a power proportional to the their mass squared, $P\sim M^2$, making the tiny ones  actually stable while large ones would quickly evaporate off. Such a scenario suggests the existence of a new kind of black hole remnants, called ``dimensional remnants'', that might generate detectable electroweak bursts \cite{Mur12}. In other words, remnants of dilaton gravity black holes might still be observable today, provided the effective ($1+1$) dimensional nucleation  survives the inflation.

\begin{figure}[ht!] 
\includegraphics[width=0.49 \textwidth]{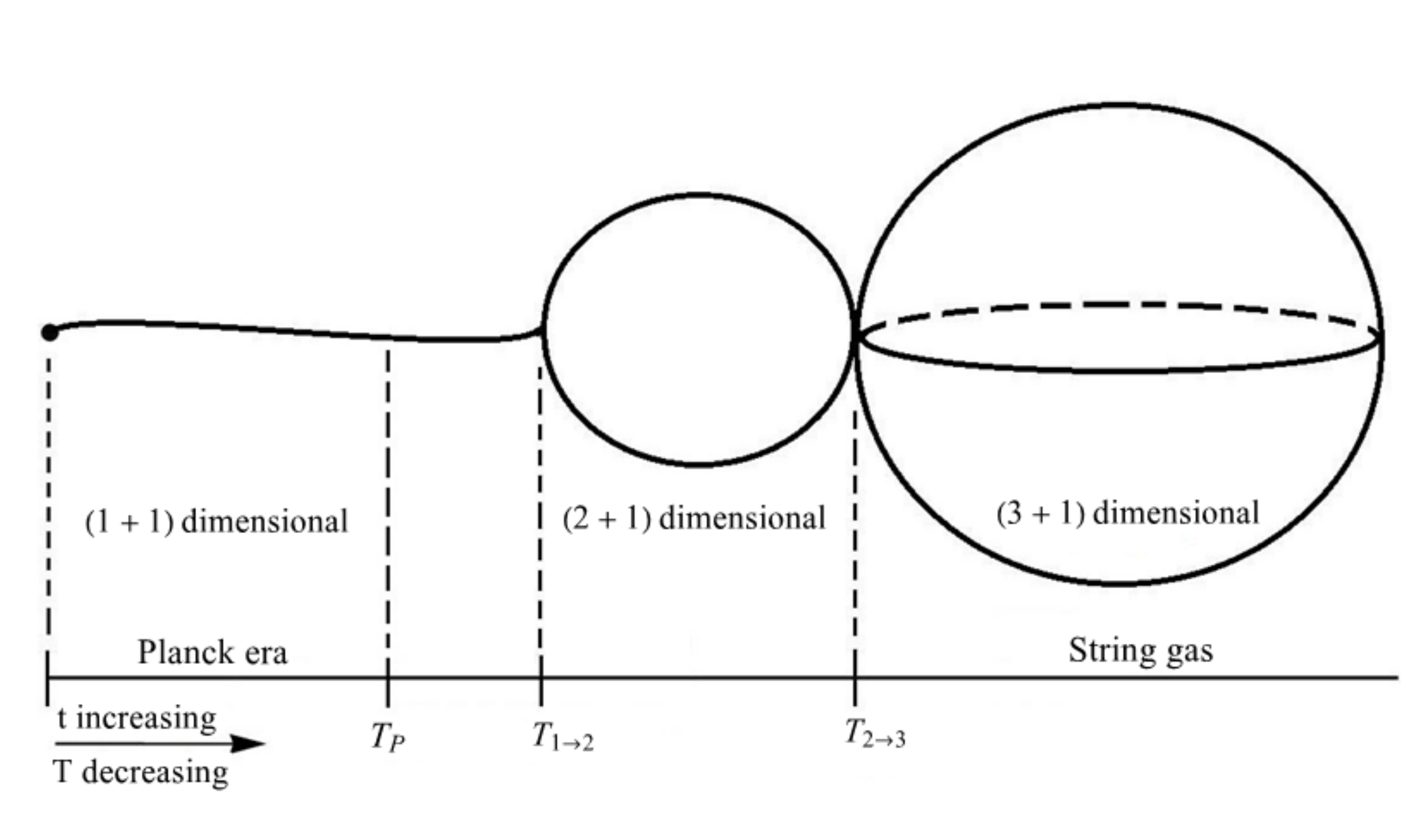}
\caption{Dimensional oxidation of the Universe. Initially, i.e., during the Planck era, the spacetime behaves as a ($1+1$) dimensional manifold; then below some transition temperature $T_{1\to 2}$, e.g.,  $T_{1\to 2}\sim T_{\text{Hag}}\sim 10^{30}\text{K} \ll T_{\text{P}}\sim 1.42 \times 10^{32}\text{K}$  (Hagedorn string transition \cite{AtW88}), it becomes ($2+1$); finally, below a critical temperature $T_{2\to 3} 
$ it takes its current conventional form. 
}
\label{fig:Fig1}
\end{figure}

 Given this background, one might be led to investigate the expansion of the early Universe, soon after the Planck era, as a process of dimensional increase, namely from an initial ($1+1$) dimensional phase to a ($3+1$) dimensional one.
 We term such a dimensional evolution  ``dimensional oxidation'', being the opposite mechanism of the reduction, as can be seen in Fig.~\ref{fig:Fig1}.
As the Universe oxidizes from ($1+1$) to ($2+1$) dimensions at a temperature $T_{1\to 2}$,  and finally to the known ($3+1$) dimensional spacetime below a temperature $T_{2\to 3}$, the existence of black holes may not be supported by pure general relativity arguments.
In ($1+1$) dimensions, gravity is necessarily described by invoking the presence of the dilaton, a scalar field $\psi$  coupled to geometry representing an extra gravitational degree of freedom alongside the graviton. The inclusion of the dilaton is crucial due to the triviality of the Einstein tensor in ($1+1$) dimensions. It should be emphasized that the dilaton action describing the ($1+1$) phase can been derived from the $D \rightarrow 2$ limit of the $D$-dimensional Einstein-Hilbert action \cite{MaR93}. This suggests  a high degree of correspondence between the  full ($3+1$) dimensional Universe and its reduced counterpart.

As the Universe oxidizes from ($1+1$) to ($2+1$) dimensions, black holes would not form in an Einstein gravity Universe, unless in the presence of an anti-de Sitter (AdS) background. For this reason, the ($2+1$) dimensional Universe may be thought   as a stage of  a non-analytic phase transition from ($1+1$) to ($3+1$) dimensional black holes \cite{TNM++18}. But what if the ($2+1$) dimensional phase is not necessarily described by Einstein's gravity? What if the dilaton and/or additional coupling survived and remained  coupled to geometry when passing from the ($1+1$) to the ($2+1$) and, finally to the ($3+1$) phase? 
Will this have an impact on PBH production?  
It is the purpose of this work to give an insight to those questions by providing a possible scenario of PBH nucleation inside an early oxidizing Universe. 

The structure of the paper is as follows: in Sec.~\ref{sec:2Dgrav} we review the ($1+1$) dimensional phase of the Universe which is governed by Liouville gravity, i.e., a linear coupling of the dilaton to gravity. 
We calculate the gravitational instantons both for the black hole and for the background (de Sitter), using the no-boundary proposal 
and we compute the pair production rate. In Sec.~\ref{sec:3Dgrav} we assume a transition to the ($2+1$) dimensional phase. We study a variety of models of dilaton gravity as well as the inclusion of a quadratic coupling to find black hole solutions. We then calculate the associated instantons and finally the rate. In Sec.~\ref{sec:4Dgrav}, we consider the  ($3+1$) dimensional case and modifications of the Einstein-Hilbert action that might have been inherited, at least in part, from the dimensionally reduced phases. As a result, we choose to work within a theory described by a part of the Horndeski action, presenting some non-trivial and regular solutions for the dilaton around a Schwarzschild-de Sitter black hole. We also calculate the pair production rate of PBHs through their instantons. Last but not least, in Sec.~\ref{sec:rates} we compare the different expressions of the rates in all three phases of the oxidizing Universe. In Sec.~\ref{sec:conclusion} we draw our conclusions. Throughout the paper we work in natural units where $c=\hbar = k_{\mathrm{B}} = 1\,$.

\section{The (1+1) dimensional phase}
\label{sec:2Dgrav}

As a start we briefly review the results found in \cite{TNM++18} for the ($1+1$) dimensional Universe. Customarily, a faithful description of a spacetime  in ($1+1$) dimensions is obtained by using  a dilatonic gravitational action, since the dilaton represents a physical connection between lower and higher dimensional spacetimes. 
In particular, Mann \& Ross showed that, in the limit $D \rightarrow 2$, the $D$-dimensional Einstein-Hilbert action in natural units reads   \cite{MaR93}:
\begin{equation} \label{2D_action}
\mathcal{S}^{(2D)} = \frac{1}{16\pi G_2} \int \mathrm{d}^2x \sqrt{-g} \Big[ \psi R + \frac{1}{2} (\nabla \psi)^2 - 2 \Lambda_{2} \Big] \,,
\end{equation}
where $R$ is the Ricci scalar, $g$ is the determinant of the metric tensor, $\psi$ is the aforementioned dilaton field and $\Lambda_2,\,G_2$ are the cosmological constant and the Newton's constant in ($1+1$) dimensions. The presence of the dilaton guarantees that the above action is a good starting point to study the dimensional oxidation.
Interestingly, while the limit $D\to 2$ has been carefully studied, the opposite one, $2 \to D$, has never been proved to be unique. This means that there can be more theories in higher dimensions that have the same $(1+1)$ dimensional limit. The Einstein-Hilbert action, from this perspective, would turn to be the leading  $(3+1)$ dimensional term in the action emerging from the process of oxidation.  This fact will be instrumental for our discussion.

Variation of the action \eqref{2D_action} with respect to the metric and the dilaton, respectively, gives
\begin{align}
\Lambda _2 g_{\mu\nu} + \frac{1}{2}\partial _{\mu} \psi \partial _\nu \psi - \frac{1}{4}&g_{\mu\nu} (\nabla \psi)^2+\nonumber \\
&+g_{\mu\nu}\Box \psi - \nabla _\mu \nabla _\nu \psi = 0 \,,\label{eq_metric}
\end{align}
\begin{equation}\label{eq_scalar}
R = \Box \psi\,.
\end{equation}
By combining the trace of \eqref{eq_metric} with \eqref{eq_scalar}, we end up with the Liouville equation in vacuum 
\begin{equation} \label{Liouv_eq}
R + 2\Lambda_2 = 0 \,,
\end{equation}
that is considered the best analogue of Einstein gravity in ($1+1$) dimensions. Assuming a linearly-symmetric solution around the origin ($x \rightarrow |x|$) and solving \eqref{Liouv_eq} for a line element of the form
\begin{equation}
\mathrm{d} s^2 = - f(|x|) d t ^2 + \frac{\mathrm{d} x^2}{f(|x|)} \,,
\end{equation}
we get the solution \cite{MST90}
\begin{equation} \label{2d_metric}
f(|x|) = C + 2M |x| + \Lambda_2 x^2 \,,
\end{equation}
where $C$ and $M$ are integration constants. Eq.~\eqref{2d_metric} represents a ($1+1$) dimensional Schwarzschild-(anti-)de Sitter black hole with mass $M$ for negative $C\,$. In such a case,  the integration constant can be normalized, $C=-1$, for sake of clarity. The sign convention for $\Lambda_2 = \pm |\Lambda_2| $ is de Sitter ($ - $) and anti-de Sitter ($+$). Also the black hole mass scales as $M \sim x^{-1}\,$, making indistinguishable the concept of a particle and a black hole in ($1+1$) dimensions \cite{MuN13}.

The decay of de Sitter space can be studied by considering gravitational instantons \cite{Haw77}, namely saddle points of the Euclidean  gravitational action  (see \cite{MaR95,BoH96} for more details).
The related pair creation rate for  black holes can be obtained in the context of the no boundary proposal \cite{HaH83}, which states that the wavefunction $\Psi$ of a quantum Universe can be approximated by
\begin{equation}
\Psi \approx e ^{-\mathcal{I}} \,,
\end{equation}
where $\mathcal{I}$ is the instanton-action. The square of this wavefunction provides the probability density to nucleate a specific Universe. Such a probability  has to be normalized by the probability of decaying de Sitter space. Therefore, the black hole pair creation rate is defined as the probability ratio of two Universes: the probability  of a Universe with a black hole $P_{\mathrm{bh}}\,$, over the probability of the empty background $P_{\mathrm{bg}}$ \cite{MaN11}:
\begin{equation} \label{rate}
\Gamma =\frac{P_{\mathrm{bh}}}{P_{\mathrm{bg}}} = \frac{e^{-2 \mathcal{I}_{\mathrm{bh}}}}{e^{-2 I_{\mathrm{bg}}}} \,.
\end{equation}
In other words, the above ``rate'' can be thought as a relative probability of two Universes and indicates how much suppressed or favored is the black hole formation.
From the relations above, one can obtain the actual probabilities as a function of the rate:
\begin{equation}
P_{\mathrm{bg}} = \frac{|\Psi_{\rm bg}|^2}{|\Psi_{\rm bg}|^2+|\Psi_{\rm bh}|^2}= \frac{1}{1+\Gamma}
\end{equation}
and
\begin{equation} \label{norm_bh_prob}
P_{\mathrm{bh}} = \frac{|\Psi_{\rm bh}|^2}{|\Psi_{\rm bg}|^2+|\Psi_{\rm bh}|^2} = \frac{\Gamma}{1+\Gamma} \,.
\end{equation}

For the ($1+1$) dimensional Schwarzschild-de Sitter black hole we get two different cases; a lukewarm black hole $\mathrm{for} \quad M > \sqrt{|\Lambda_2|}$ \cite{Rom92} 
and a Nariai black hole \cite{Bou97} with the event and the cosmological horizon coalescing in a degenerate horizon ($x_{\mathrm{h}} = x_{\mathrm{c}} = 1 / \sqrt{|\Lambda_2|}=1/M$). 

The instanton-action can be derived from the Wick-rotated version of \eqref{2D_action}, namely
\begin{equation} \label{2d_action-insta}
\mathcal{I}^{(2D)}=  - \frac{|\Lambda_2|}{8 \pi G_{2} } \int \mathrm{d} ^2x \sqrt{g} \     \left(   \frac{1}{2} \psi  +1  \right) \, .
\end{equation}
To get the above compact form of the action, one has to perform an integration by parts of the instanton kinetic term and  use the equations of motion of Liouville gravity. The dilaton resulting from \eqref{eq_metric} and \eqref{eq_scalar} has the form of
\begin{equation}
\psi(t,x) = \psi_0 + \frac{4\pi}{\beta} t -\ln f(|x|) \,, 
\end{equation}
where $\psi_0$ is an integration constant and $\beta$ is the periodicity of the Euclidean time given by the inverse of the Hawking temperature. The $t$-dependence of the dilaton is necessary for its regularization on the horizons, a feature that can be verified by making a coordinate transformation to the Eddington-Finkelstein coordinate system.

After choosing the suitable boundary $\psi_0$  for the dilaton, we find that the instanton of the de Sitter background depends on integration constants only. Therefore it can be set to zero, $\mathcal{I}^{(2D)}_{\mathrm{dS}} = 0$.
This means that de Sitter background will not affect the production rate. In contrast to standard general relativity in ($3+1$) dimensions, the quantum instability of de Sitter space occurs even if  the cosmological constant does not attain Planckian values. This is the result of the absence of a fundamental energy scale in the coupling constant, $G_2$, of ($1+1$) dimensional gravity.

After calculating the instantons for the black hole cases, one can finally get the production rates.
For the lukewarm black hole the production rate reads:
\begin{equation} \label{2d_luke_rate}
\Gamma^{(1+1)}_{\mathrm{lw}}=  \left( \frac{1}{ M^2 / |\Lambda_2| - 1} \right)^{1/2G_{2}} \quad \mathrm{with} \quad M > \sqrt{|\Lambda_2|} \,. 
\end{equation} 
In marked contrast to the ($3+1$) case \cite{BoH95}, this rate is not exponentially suppressed. Specifically, one finds:
\begin{itemize}
\item for $M \approx \sqrt{|\Lambda_2|} \,$, the rate diverges ($\Gamma_{\mathrm{lw}} \gg 1$);\item   for  $\sqrt{|\Lambda_2|} < M < \sqrt{2|\Lambda_2|} \,$, the rate $\Gamma_{\mathrm{lw}}$ exceeds unity, 
corresponding to a highly unstable de Sitter space;  
\item for $M=\sqrt{2 |\Lambda_2|} \,$,   $\Gamma_{\mathrm{lw}}=1 \,$, meaning that the two Universes have equal probabilities.
\end{itemize}
Only for $M \gg \sqrt{|\Lambda_2|} \,$, the rate is highly suppressed $(\Gamma_{\mathrm{lw}} \ll 1) \,$. This corresponds to the case in which the black hole mass is larger than the energy stored in the cosmological term. As a result the decay is not possible.

For the Nariai black hole the rate has to depend on an additional mass scale $\mu_0$, which is arbitrary and not related to $M$ or to the Planck mass $M_{\mathrm{P}}\,$. This occurs because, in contrast to the lukewarm case, the Nariai rate cannot depend on the ratio $M^2 / |\Lambda_2|$, being $M=\sqrt{|\Lambda_2|}$ the condition for the existence of the instanton.  
As a result one finds:
\begin{equation} 
\Gamma_{\mathrm{N}}^{(1+1)} = \left( \frac{ \mu _0 ^2}{|\Lambda_2 |} \right)^{1/2G_2}  \, .
\label{2d_nariair_ate}
\end{equation}
Being $\mu_0$ not set {\it a priori}, Nariai PBHs can be prolifically produced  for any value of $\Lambda_2$ even in the sub-Planckian mass regime.

\section{The (2+1) dimensional phase}
\label{sec:3Dgrav}

As we already mentioned, in ($1+1$) dimensions there is no possibility for gravity to be described solely by the Einstein-Hilbert action and that is why we need the introduction of the dilaton.

Works on gravity in ($2+1$) dimensions date back to more than forty years ago \cite{CLEMENT1976437,Collas2+1}, but the topic became of major interest only later  \cite{Deser:1983tn,Achucarro:1987vz,Witten:1988hc}. It turns out that, general relativity in ($2+1$) dimensions has no propagating degrees of freedom, or in other words, there are no gravitons in its quantum description. This happens because of the vanishing of the Weyl tensor and so the remaining curvature tensor is described by the Ricci tensor and scalar. In fact, the spacetime will be locally flat, de Sitter or anti-de Sitter in vacuum, depending on the value of the cosmological constant. That is why it is sometimes called ($2+1$) topological gravity. 

For many years it was thought that, the vanishing of the ($2+1$) Riemann tensor in the absence of a cosmological term would lead to non-existence of black holes. 
However, Ba\~nados, Teitelboim and Zanelli found the so-called BTZ black hole \cite{Banados:1992wn} in the presence of a negative cosmological constant, which resembles the properties of the Schwarzschild and Kerr black holes in ($3+1$) dimensions.

Along this line of reasoning, the derivation of ($2+1$) black hole solutions embedded in a de Sitter space have required the inclusion of higher derivative terms in the gravity action. Let us consider, for example, the action of Bergshoeff-Hohm-Troncoso (BHT) massive gravity \cite{BHT09}
\begin{equation} \label{bht_action}
\mathcal{S}^{(3D)} = \frac{1}{16\pi G_3} \int \mathrm{d}^3x \sqrt{-g} \left( R - 2\Lambda _3 - \frac{1}{m^2} K \right) \,,
\end{equation}
where $K= R_{\mu\nu} R^{\mu\nu} - \frac{3}{8} R^2$ and $m$ is a mass parameter. This theory predicts massive gravitons with two spin states of helicity $\pm 2$. By varying the action \eqref{bht_action} with respect to the
metric we get
\begin{equation}
G_{\mu\nu} + \Lambda_3 g_{\mu\nu} - \frac{1}{2m^2} K_{\mu\nu} = 0 \,,
\end{equation}
with
\begin{align} \nonumber
K_{\mu\nu} = &2 \square R_{\mu\nu} - \frac{1}{2} \nabla _{\mu} \nabla _{\nu} R - \frac{1}{2} g_{\mu\nu} \square R - 8 R_{\mu \lambda} R^{\lambda}{}_{\nu}+ \\
 &+ \frac{9}{2} R R_{\mu\nu} + g_{\mu\nu} \left( 3R^{\kappa \lambda} R_{\kappa \lambda} - \frac{13}{8} R^2 \right) \,.
\end{align}
It has been proven \cite{OTT09,Mae11} that for the special case of $\Lambda_3 =m^2\,$, the theory admits a unique spherically symmetric solution. Since we are interested in a Schwarzschild de Sitter like solution, we consider a static and spherically symmetric line element of the form
\begin{equation}
ds ^2 = - f(r) dt^2 + f(r)^{-1}dr^2 + r^2 d\phi ^2\,.
\end{equation}
 with metric potential
\begin{equation}\label{bh_in2+1D}
f(r) = -G_3 \mu + b r - \Lambda_3 r^2 \,,
\end{equation}
where $\mu$ is a black hole mass parameter and $b$ is an integration constant scaling as $b \sim r^{-1}$, that is non-vanishing for $1/m^2\neq 0$. One sees that black holes exist not only for $\Lambda_3 < 0\,$,  but also for $\Lambda_3 > 0$ where we can get two different horizons, i.e., one event horizon $r_+$ and one cosmological horizon $r_{\mathrm{c}}\,$. If $b = 0$, there is no contribution from the higher derivative term $K$, then for $\Lambda _3 <0 $ the solutions becomes the known BTZ black hole \cite{BTZ92}. 
From this perspective we can already anticipate that the ($2+1$) phase can be either a non-analytic phase of the oxidation without black holes or a smooth transient phase admitting event horizons. To better understand how this phase connects with the preceding ($1+1$) dimensional era, let us see what is the role of the dilaton in ($2+1$) dimensional gravity,  without considering higher derivative terms. 

We start from the general action
\begin{equation}\label{2+1D_action}
\mathcal{S}^{(3D)} = \frac{1}{16\pi G_3} \int \mathrm{d}^3x \sqrt{-g} \left(h(\psi) R -\omega \left(\nabla  \psi\right)^2 - 2V(\psi) \right) \,,
\end{equation}
where $h(\psi)$ is a function of $\psi$, determining whether the coupling of the dilaton will be minimal ($h(\psi) = 1$) or not, $V(\psi)$ is an arbitrary potential and $\omega$ is a coupling constant. If the potential is not dynamical, it is straightforwardly associated with the cosmological constant in ($2+1$) dimensions $\Lambda _3$. By varying the action \eqref{2+1D_action} with respect to the metric and the scalar field respectively, we get
\begin{align}
h(\psi) G_{\mu\nu} + V(\psi) g_{\mu \nu} = &\omega \left(\nabla _\mu\psi \nabla _\nu \psi - \frac{1}{2}g_{\mu\nu}(\nabla \psi)^2\right) -\nonumber\\
&- g_{\mu\nu}\Box h + \nabla _\mu \nabla _\nu h\,,\\
2 \omega \Box \psi + \frac{dh}{d\psi}R = &2 \frac{dV}{d\psi}\,.
\end{align}

For several different theories, with a variety of coupling functions $h(\psi)$ and potentials $V(\psi)$, one can see that black holes form in the presence of an anti-de Sitter background only (see Tab. \ref{tab:1}). This would confirm the fact that, a violation of the dominant energy condition is required for the existence of de Sitter black holes in ($2+1$) dimensions \cite{Ida:2000jh}.

Other examples known in the literature confirm such a scenario. Minimally or non-minimally coupled scalar fields have been considered in ($2+1$) dimensions, but existence of  black holes always necessitates AdS asymptotics \cite{Henneaux:2002wm,Fan:2015tua,Martinez:1996gn,AyonBeato:2004ig,AyonBeato:2001sb,
Xu:2013nia,Zhao:2013isa,Tang:2019jkn}.

\begin{table}[!h]
\begin{tabular}{c|c|c}
$h(\psi)$ & $\omega$ & $V(\psi)$ \\
\hline
\hline
$\psi$, $\psi^2$ & $1/2$ & $\Lambda _3$, $\Lambda _3 + V_0 \psi^2$, $\Lambda _3 + V_0 \psi^4$, $\Lambda _3 + V_0 \psi^6$\\
\hline
$1-\psi^2/8$ & $1/2$ & $\Lambda _3$, $\Lambda _3 + V_0 \psi^2$, $\Lambda _3 + V_0 \psi^4$, $\Lambda _3 + V_0 \psi^6$\\
\end{tabular}
\caption{Theories under consideration in ($2+1$) dimensions, all requiring negative cosmological term for black hole existence.}
\label{tab:1}
\end{table}

Given this background we can briefly summarize what one can learn from the proposed analysis up to now. In ($1+1$) dimensions, gravity is necessarily mediated through the metric and the scalar field. In ($3+1$) dimensions, there are many reasons (see e.g. \cite{Clifton:2011jh,Capozziello:2011et,Nojiri:2017ncd}) to believe that general relativity is not the final theory of gravity to successfully describe the cosmological dynamics. During the intermediate phase, the ($2+1$) the scalar field trivializes and one has to introduce higher derivative terms to find black holes with de Sitter asymptotics. This means that the theory governing the gravitational interactions in this phase would be of the form 
\begin{align}
\mathcal{S}^{(3D)} = \frac{1}{16\pi G_3}\int d^3 x \sqrt{-g}\Big(R - 2 \Lambda _3 &- \frac{1}{2} (\nabla \psi)^2 \nonumber \\
&- \frac{1}{m^2}K\Big)\,, 
\end{align}
where the dilaton couples minimally to the metric, and its solution is constant. In other words, the above action is actually equivalent to  \eqref{bht_action}. Such a possibility would suggest that the ($3+1$) dimensional Universe could have inherited correcting terms to the gravity action from the preceding dimensionally reduced phases, i.e., the dilaton as well as higher derivative terms. This fact will be instrumental to calculate the black hole production rate in ($3+1$) dimensions -- see the following section.

After this premise, we will proceed by calculating the pair creation for the theory \eqref{bht_action}.

\subsection*{Pair Creation Rate in BHT massive gravity}
\label{subsec:BHT_gravity}

Along the lines of what found in \cite{OTT09}, we start by calculating the pair creation rate of Nariai black holes  in BHT massive gravity \eqref{bht_action}. Taking the condition $f(r)=0$ along with $\Lambda_3 > 0\,$ from \eqref{bh_in2+1D}, we find that
\begin{align}
r_+ &= \frac{1}{2\Lambda_3} \left(  b- \sqrt{b^2 - 4 \mu G_3 \Lambda_3} \right) \,, \\
r_{\mathrm{c}} &= \frac{1}{2\Lambda_3} \left(  b + \sqrt{b^2 - 4 \mu G_3 \Lambda_3} \right) \,.
\end{align}
At the special case of $\mu = \frac{b^2}{4G_3 \Lambda_3}\,$, the two horizons coincide at $\rho = r_+ = r_{\mathrm{c}}=b/2\Lambda _3\,$. Therefore, a black hole exists for the mass range $0 < \mu < \frac{b^2}{4G_3 \Lambda_3} \,$. 

Remarkably, the two horizons have always a common Hawking temperature $T_+ = T_{\mathrm{c}}= \frac{1}{4\pi} \sqrt{b^2 - 4 \mu G_3 \Lambda_3} $ and in the Nariai limit, $r_+ = r_{\mathrm{c}}$,  this temperature vanishes. After Wick-rotating ($\tau = it$), we can build a regular instanton,  whose line element    can be expressed with respect to the two horizons. Taking into account that
\begin{equation}
\mu = \frac{\Lambda_3 r_+ r_{\mathrm{c}}}{G_3} \qquad \mathrm{and} \qquad b = \Lambda_3 (r_+ + r_{\mathrm{c}}) \,,
\end{equation}
the form of $f(r)$ is given by $f(r) = \Lambda_3 (r- r_+) (r_{\mathrm{c}} -r)$  and the black hole instanton reads
\begin{equation} 
\mathrm{d} s^2 =  \Lambda_3 (r- r_+) (r_{\mathrm{c}} -r) \mathrm{d} \tau^2 + \frac{\mathrm{d} r^2}{\Lambda_3 (r- r_+) (r_{\mathrm{c}} -r)}  + r^2 \mathrm{d} \phi ^2 \,.
\end{equation}
For the Nariai case we can apply a transformation of the form \cite{Bou97}
\begin{equation}
\tau = \frac{\xi}{\epsilon \Lambda _3}\,, \quad r = \rho - \epsilon \cos \chi\,, 
\end{equation} 
where $\xi$ and $\chi$ are periodic variables with periods $2\pi$ and $\pi$ respectively. The length $\epsilon$ is an arbitrarily small scale ($\epsilon \ll G_{3}$) with $\epsilon \rightarrow 0$ in the Nariai limit. That means that the event horizon lies at $r_{\rm +} = \rho - \epsilon$ and the cosmological one at $r_{\rm c}=\rho + \epsilon\,$. Then the metric potential takes the approximate form of $f(r) \approx \Lambda _3 \epsilon ^2 \sin ^2 \chi$ and the Nariai instanton reads
\begin{equation}
\mathrm{d} s^2 = \frac{1}{\Lambda_{3}} \left(  \mathrm{d} \chi ^2 + \sin ^2 \chi \ \mathrm{d} \xi ^2 + \frac{b^2}{4\Lambda_3} \mathrm{d} \phi ^2 \right) \,.
\end{equation}
Now we can estimate the pair creation rate  by calculating the two instanton-actions from the Wick-rotated version of the BHT-action \eqref{bht_action}. The evaluation of the instanton-action vanishes for the above two black hole solutions 
\begin{equation}
\mathcal{I}_{\mathrm{bh}}^{(3D)} = \mathcal{I}^{(3D)}_{\mathrm{Nariai}} = 0 \,,
\label{eq:3dobjinst}
\end{equation}
while the instanton action of the respective de Sitter background is
\begin{equation}
\mathcal{I} _{\mathrm{bg}}^{(3D)} = - \frac{\pi}{2 G_3 \sqrt{\Lambda_3}} 
\end{equation}
and so the pair creation rate \eqref{rate} becomes
\begin{equation}
\Gamma_{(\mathrm{BHT})} = e^{-\frac{\pi}{G_3 \sqrt{\Lambda _3}}} \,.
\end{equation}
Effectively, this means that we can retrieve an expanding de Sitter Universe whose spontaneous black hole formation is suppressed relative to the empty de Sitter background. We also note that, despite both instantons \eqref{eq:3dobjinst} vanish, the above rate does implicitly depends on the black hole parameters through the cosmological constant $\Lambda _3$.

\section{The (3+1) dimensional phase}
\label{sec:4Dgrav}

For almost two decades, there have been many attempts \cite{Clifton:2011jh,Capozziello:2011et,Nojiri:2017ncd} to find a better description for gravity than general relativity, since it appears that Einstein's theory is plagued with many shortcomings both in the short and in large scale regime. 
Scalar-tensor theories  have been among the first proposals  alternatives to general relativity. Already in the early sixties Brans and Dicke conjectured the presence of a scalar field non minimally coupled to gravity. However, the most general scalar-tensor theory in four dimensions, with a single scalar field, leading to second order field equations was firstly proposed by G.~W.~Horndeski \cite{Horndeski1974}, in 1974, and later rediscovered as the decoupling limit of the five dimensional DGP massive gravity \cite{Nicolis:2008in,Deffayet:2009wt}.

It is well known though that, once a black hole reaches a stationary state, it is characterized only by its mass, its charges (i.e. associated with long range gauge fields) and its angular momentum. This is the so-called no-hair theorem, that has also been extended in the Brans-Dicke \cite{Hawking1972,Bhattacharya:2015iha} and other scalar-tensor theories \cite{Sotiriou:2011dz}. In these papers it has been shown that there is no non-trivial and regular solution for the scalar field around a black hole, or shortly, black holes have no scalar hair. Conversely, if we allow higher order derivative couplings in the action (while the equations are still of second order), there are non-trivial and regular solutions for the scalar field around a black hole.

Armed with the results from previous sections, we are ready to exploit the dimensional oxidation to select a suitable action containing both a scalar field and higher order derivative terms. 
 This is why in the present section we consider a theory that is part of the Horndeski action. Such a set up would allow for the study of black hole production that can potentially depart from the previous  Bousso Hawking result about the known Schwarzschild de Sitter spacetime \cite{BoH95}.

\subsection{Galileon black holes}

Among the theories described by the Horndeski action we are considering  those that are shift symmetric under the transformation $\psi \rightarrow \psi + \text{constant}$. The advantage of this choice is that  such theories are customarily taken in account in cosmological contexts since (as we will show) they offer self-accelerating solutions at late-times. Specifically, the chosen action reads
\begin{align} \nonumber
\mathcal{S}^{(4D)} = \frac{1}{16\pi G_{\mathrm{N}}} \int d^4x &\sqrt{-g} \Big[  R - \omega ( \nabla \psi)^2 +\\ \label{galileon_action}
 &+  B G_{\mu\nu} \nabla^{\mu}\psi \nabla^{\nu}\psi  - 2\Lambda \Big]\,,
\end{align}
where $G_{\mathrm{N}}$ is the known Newton's constant in four-dimensions,  $\Lambda$ is the cosmological constant and $\omega,\,B$ are coupling constants \cite{Babichev:2013cya}. 
Apart from the obvious $\omega >0\,$, in order to have no ghost instabilities for the scalar field, we also have to consider $B<0$  to get de Sitter-like solutions, as we will see later on. 
As anticipated from above, the presence of the third term in the action, i.e., the ``John'' term \cite{Charmousis:2011bf},  which denotes the non-trivial interaction of the geometry with the scalar field, will help us overcome the known no-hair theorems \cite{Sotiriou:2011dz,Bhattacharya:2015iha} and will give non-trivial solutions for the dilaton in a static and spherically symmetric black hole geometry. The John term contains the higher derivative terms we assumed to be inherited from the ($2+1$) dimensional phase, even if they exhibit a different functional dependence.

Before we cut to the chase, let us pause to motivate a bit more the chosen model, i.e. \eqref{galileon_action}. First of all, as already mentioned, the theory is part of the Horndeski action and as such, even though it contains higher order coupling terms, it does not propagate new degrees of freedom (two of the metric plus one of the scalar field). Furthermore, this very coupling was proposed as a remedy to Higgs inflation that suffered from dangerous quantum corrections, and is known as \textit{new} Higgs infation \cite{Germani:2010gm}. Last but not least, the same higher order term appears when one takes the $D \to 4$ limit of the Gauss-Bonnet gravity \cite{Hennigar:2020lsl}.

The last few months there has been an increased interest in the so-called Einstein-Gauss-Bonnet theories of gravity in ($3+1$) dimensions, see \cite{Glavan:2019inb,Gurses:2020ofy,Ai:2020peo,Hennigar:2020fkv,Hennigar:2020zif} and references therein. It is reasonable thus to ask what if the Universe oxidized from $(1+1)$ to $(2+1)$ and finally to $(3+1)$ dimensions as the limit of the Gauss-Bonnet gravity in arbitrary $D$ dimensions. The problem however arises from the fact that in $D < 4$ the Gauss-Bonnet term identically vanishes. On top of that, in (2+1) dimensions, the Riemann tensor vanishes as well, thus surviving only terms like $K$ in \eqref{bht_action}. There has been some studies in lower dimensional Gauss-Bonnet gravity \cite{Hennigar:2020fkv,Hennigar:2020zif}, where they find BTZ like black holes, however, all of them have anti-de Sitter asymptotics. 
Proceeding now with the variation of the action \eqref{galileon_action} with respect to the metric we get
\begin{align}
 G_{\mu\nu}  +\Lambda g_{\mu\nu} = &\omega \left(\nabla_{\mu}\psi\nabla_{\nu}\psi-\frac{1}{2}g_{\mu\nu}(\nabla \psi)^2 \right) -\nonumber \\
&- \frac{B}{2}\Big(G_{\mu\nu}(\nabla \psi)^2 + 2 P_{\mu\alpha\nu\beta}\nabla^{\alpha}\psi\nabla^{\beta}\psi +\nonumber \\
& +g_{\mu\alpha}\delta ^{\alpha\rho\sigma}_{\nu\gamma\delta}\nabla^{\gamma}\nabla_{\rho}\psi\nabla^{\delta}\nabla_{\sigma}\psi \Big) \,,\label{met_eom_galileon}
\end{align}
where $P_{\alpha\beta\mu\nu}$ is the dual of the Riemann tensor, 
\begin{equation}
P_{\alpha\beta\mu\nu} = -\frac{1}{4}\epsilon _{\alpha\beta\rho\sigma}R^{\rho\sigma\gamma\delta} \epsilon_{\mu\nu\gamma\delta}\,.
\end{equation}
One can easily notice that for $\omega = 0 = B$, we can recover Einstein equations with a cosmological constant. The related black hole solutions is the Schwarzschild-de Sitter geometry. Moreover, if we vary the action with respect to the dilaton, we get the equation of motion for it, that can be written as:
\begin{equation}\label{scal_eom_galileon}
\nabla_{\mu}J^{\mu} = 0 \,, \quad \text{with}\quad J^{\mu} = \left(\omega g^{\mu\nu} - B G^{\mu\nu}\right) \nabla_{\nu}\psi \,.
\end{equation}
A detailed discussion about the behavior of the scalar field itself, or of the induced Noether current at the horizon, can be found in \cite{Babichev:2013cya}. Rather, we will directly proceed to the scalar solution around a self-tuning Schwarzschild-de Sitter black hole, which is of interest in our paper.

We start by considering a static and spherically symmetric metric of the form\footnote{Obviously, it is not the most general one, but we are interested in metrics of Schwarzschild-de Sitter type.}
\begin{equation}
\mathrm{d} s^2 = - f(r) \mathrm{d} t^2 + f(r)^{-1} \mathrm{d} r^2 + r^2 \mathrm{d} \Omega ^2\,,
\end{equation}
with $\mathrm{d} \Omega ^2 = \mathrm{d} \theta ^2 + \sin ^2 \theta \ \mathrm{d} \phi ^2$ and a scalar field which has, apart from the $r$ dependence from the metric, an additional time-dependence, i.e., $\psi = \psi (t,r)\,$. 
It has been shown in \cite{Babichev:2013cya}  that equations  \eqref{met_eom_galileon} and  \eqref{scal_eom_galileon} are satisfied when
\begin{gather}
f(r) = 1 - \frac{2\mu}{r} - \frac{\Lambda _{\rm eff}}{3}r^2  \,, \label{met_sol}\\
\psi(r) = q t + \phi(r)\,, \label{gal_sol}
\end{gather}
with
\begin{equation} \label{3+1_dil}
\Lambda_{\rm eff}= \omega / |B|  \quad \mathrm{and} \quad  \phi'(r) = \pm \frac{q}{f}\sqrt{1-f} \,.
\end{equation}
The mass parameter $\mu$ can be written as $\mu = a G_{\mathrm{N}} M$ where $M$ is  the  mass of the black hole and the positive constant $a$ can be specified from the corresponding Newtonian limit. The parameter $q$ has dimensions of an energy, $[E]$, and its value is specified by $q^2 \omega = \Lambda- \Lambda _{\rm eff}$. 
From this relation we see that $ \Lambda \geq \Lambda_{\text{eff}} $ must be satisfied for \eqref{gal_sol} to be real (we remind that $\omega >0$ as the standard kinetic term). Note that Eq.~\eqref{met_sol} can be also expressed as $f(r) = 1 - \frac{2\mu}{r} - \frac{1}{3} \left(\frac{\Lambda}{1+q^2 |B|}\right) r^2$. 

\subsection{Gravitational instantons}
We build again two instantons  from the metric \eqref{met_sol}: one for the de Sitter background and one for the Nariai black hole.

For the de Sitter instanton ($\mu=0$ in \eqref{met_sol}), the metric potential is $f_{\mathrm{dS}}(r)=1-\frac{\Lambda_{\mathrm{eff}}}{3} r^2\,$. The cosmological horizon lies at $r_{\mathrm{c}}=\sqrt{3/\Lambda_{\mathrm{eff}}}$ with a temperature of $T=\frac{1}{2\pi r_{\mathrm{c}}}=\beta ^{-1}\,$. The de Sitter instanton then reads
\begin{equation} \label{ds_instanton2}
\mathrm{d} s^2 = f_{\mathrm{dS}}(r) \mathrm{d} \tau^2 + f_{\mathrm{dS}}(r)^{-1} \mathrm{d} r^2 + r^2 \mathrm{d} \Omega ^2\,.
\end{equation}
After performing the transformation $\tau = r_{\mathrm{c}} \xi   ,\, r =r_{\mathrm{c}} \cos \chi \,$, the above instanton can be cast into the regular form of
\begin{equation} \label{ds_action2}
\mathrm{d} s^2 = \frac{3}{\Lambda_{\mathrm{eff}}} \left(  \mathrm{d} \chi ^2 + \sin ^2 \chi \ \mathrm{d} \xi ^2 +  \cos ^2 \chi \ \mathrm{d} \Omega ^2 \right) \,.
\end{equation}
In addition, since we have Wick-rotated the time axis, the dilaton \eqref{gal_sol} becomes $\psi = -i q \tau + \phi (r)\,$. 

 For the Nariai black hole we have the relations $9 \mu ^2 \Lambda_{\mathrm{eff}} = 1$ and $\rho=3\mu=1/\sqrt{\Lambda_{\mathrm{eff}}}$ where $\rho=r_2=r_3$ is the degenerate horizon. After a  transformation for approximately degenerate black holes
\begin{equation} \label{Nariai-transf2}
\tau = \frac{\xi}{\epsilon \Lambda_{\mathrm{eff}} } \,, \ \ r = \rho - \epsilon \cos \chi \,, \ \  r_2 = \rho-\epsilon  \,, \ \  r_3 = \rho + \epsilon \,,
\end{equation}
with $\epsilon $ being again a small length parameter, the metric potential \eqref{met_sol} takes the approximate form of $f_{\mathrm{N}}(\chi) \approx \Lambda_{\mathrm{eff}} \epsilon ^2 \sin ^2\chi$ and, in the limit $\epsilon \rightarrow 0\,$, the instanton reads
\begin{equation}
\mathrm{d} s^2 = \frac{1}{\Lambda_{\mathrm{eff}}} \left(  \mathrm{d} \chi ^2 + \sin ^2 \chi \ \mathrm{d} \xi ^2 +  \mathrm{d} \Omega ^2 \right) \,.
\end{equation}
Also the dilaton becomes
\begin{equation}\label{gal_sol2}
\psi = \psi(\xi,\chi) = - \frac{i q \xi}{\epsilon \Lambda_{\mathrm{eff}} } + \phi(\chi) \,,
\end{equation}
with
\begin{equation}
\frac{\partial \phi (\chi)}{\partial \chi} = \left( \frac{q}{f_{\mathrm{N}}} \sqrt{1- f_{\mathrm{N}}} \right) \frac{\partial r}{\partial \chi} \,.
\end{equation}
Note that even though the dilaton, both in \eqref{gal_sol} and in \eqref{gal_sol2}, may look divergent on the horizon or when $\epsilon \rightarrow 0$, it is not. One can verify this by changing to the generalized Eddington-Finkelstein coordinates.

\subsection{Pair creation rate}

The evaluation of the Euclidean version of the action \eqref{galileon_action} for the de Sitter and the Nariai instanton will give respectively
\begin{equation} \label{4d_ds_action}
\mathcal{I}_{\mathrm{dS}}^{(4D)} = - \frac{3\pi}{2G_{\mathrm{N}} \Lambda_{\mathrm{eff}}^2} \Big[ 2\Lambda_{\mathrm{eff}}  - \Lambda  \Big] \,
\end{equation}
and
\begin{equation} \label{4d_nariai_action}
\mathcal{I}_{\mathrm{N}}^{(4D)} = - \frac{\pi}{G_{\mathrm{N}} \Lambda_{\mathrm{eff}}^2 } \Big[  2 \Lambda_{\mathrm{eff}} - \Lambda  \Big] \,.
\end{equation} 
A note may be necessary here: even though the bare cosmological constant $\Lambda$ does not appear in the spacetime metric, it appears explicitly in the solution of the scalar field, through $q$ and also in the action \eqref{galileon_action}; that is why it enters the instanton action. Substituting \eqref{4d_ds_action} and \eqref{4d_nariai_action} in \eqref{rate}, we get a Nariai rate of the form
\begin{align} \nonumber
\Gamma^{(3+1)}_{\mathrm{N}} &= \exp \Big[ - \frac{\pi}{G_{\mathrm{N}} \Lambda_{\mathrm{eff}}^2} \Big( 2\Lambda_{\mathrm{eff}}  - \Lambda \Big)   \Big] \\ \label{4d_rate}
&= \exp\left[  - \frac{\pi}{G_{\mathrm{N}} \Lambda} \left( 1-B^2 q^4 \right)  \right] \,.
\end{align}
We recall that, in the $|B|\rightarrow 0$ limit, there is  no non-trivial solution for the scalar field for any $\omega$. This means that black holes (as well as the de Sitter Universe)  and the above instantons  are exactly the same as in general relativity. This confirms that black holes with non-trivial dilatonic configurations can exist only in the presence of the higher derivative coupling (John term). 

We also remark that the rate \eqref{4d_rate} exceeds the conventional Bousso-Hawking (BH) rate  \cite{BoH95}: for any non-vanishing $q$ and $B$ one has $\Gamma^{(3+1)}_{\mathrm{N}} >\Gamma_\mathrm{BH} = e^{-\pi/\Lambda G_{\rm N}}\,$.   In addition the dilaton coupling provides a richer phenomenology: 
\begin{itemize}
\item For $  \Lambda > 2\Lambda_{\mathrm{eff}}$ (or $|B|>1/q^2$ ) the rate is unsuppressed signalling an overproduction of PBHs.
\item For $2\Lambda_{\mathrm{eff}}  = \Lambda$ (or $|B| = 1/q^2$ ) the two Universes have equal probabilities to nucleate.
\item For $\Lambda_{\mathrm{eff}}  < \Lambda < 2\Lambda_{\mathrm{eff}}$ (or $|B| < 1/q^2$ ) the pair production is exponentially suppressed.
\end{itemize}
This result follows a natural continuation of the expansion of the Universe. In our model both the dilaton and the cosmological constant drive  the inflationary era. At the end of inflation, there might be a time window where the value of $\Lambda$ is relatively higher compared to $2\Lambda_{\mathrm{eff}}$ ($\Lambda \geq 2\Lambda_{\mathrm{eff}}$), allowing this way the prolific production of relatively large and cold black holes with respect to the Planckian ones. As the Universe continues to expand, the value of $\Lambda$ decreases even further, reaching the range of values $\Lambda_{\mathrm{eff}}  < \Lambda < 2\Lambda_{\mathrm{eff}}$ at late times where the production stops. We remind here that $\Lambda \geq \Lambda _{\rm eff}$ always, in order for the dilaton to be real. This condition  would also save the catastrophic instability of de Sitter space in the late Universe \cite{DNT18}.

\section{Comparison of the rates}
\label{sec:rates}

Now the scenario of the oxidation has been completed. Assuming that the theories describing gravitational interactions in each phase are those described in the previous section, the Nariai rates  for each dimensional phase are
\begin{align} \label{rate1}
\Gamma_{\mathrm{N}} ^{(1+1)} &= \left(  \mu _0 / \sqrt{| \Lambda _2 |} \right) ^ {1/G_2} \,, \\ \label{rate2}
\Gamma_{\mathrm{N}} ^{(2+1)} &= \exp \left[-\frac{\pi}{G_3 \sqrt{\Lambda_3}}\right] \,, \\ \label{rate3}
\Gamma_{\mathrm{N}} ^{(3+1)} &=  \exp \Big[ - \frac{\pi}{G_{\mathrm{N}} \Lambda_{\mathrm{eff}}^2} \Big( 2\Lambda_{\mathrm{eff}}  - \Lambda \Big)   \Big]  \,.
\end{align}
As a first comment one can see that the dimensionality of the gravitational coupling constant affects the functional dependence of rates in each of the above regimes.

\begin{figure}[ht!] 
\includegraphics[width=0.45 \textwidth]{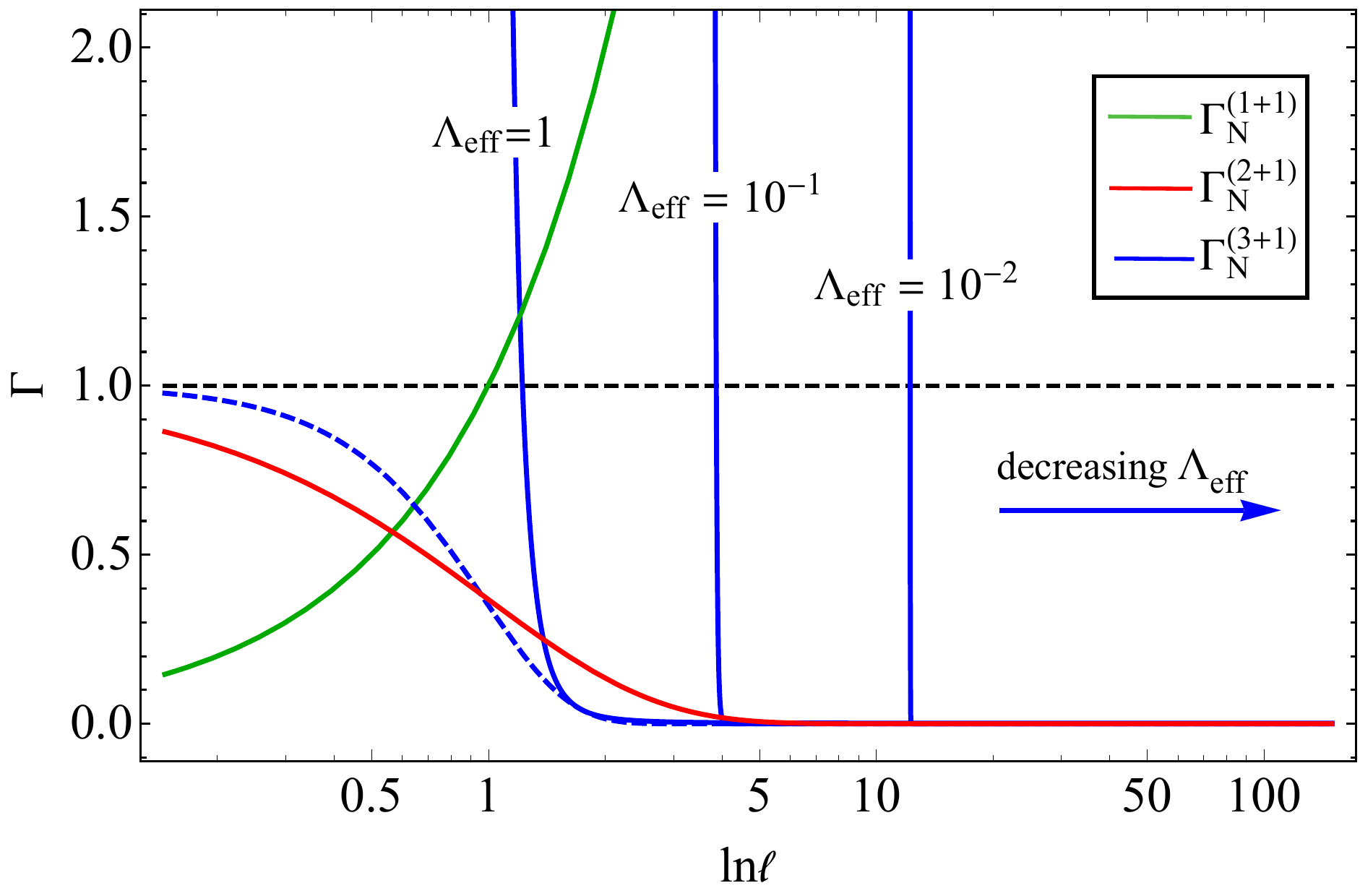}
\caption{The Nariai rates \textit{vs} the logarithm of the de Sitter radius $\ell$ are plotted for $|\Lambda_2| = \Lambda _3 = \Lambda$ and for $G_{2}=G_{3}=G_{\mathrm{N}}=\mu_0=1\,$. The blue dashed line stands for the conventional Bousso-Hawking rate.} 
\label{fig:Fig2}
\end{figure}

\begin{figure}[ht!] 
\includegraphics[width=0.47 \textwidth]{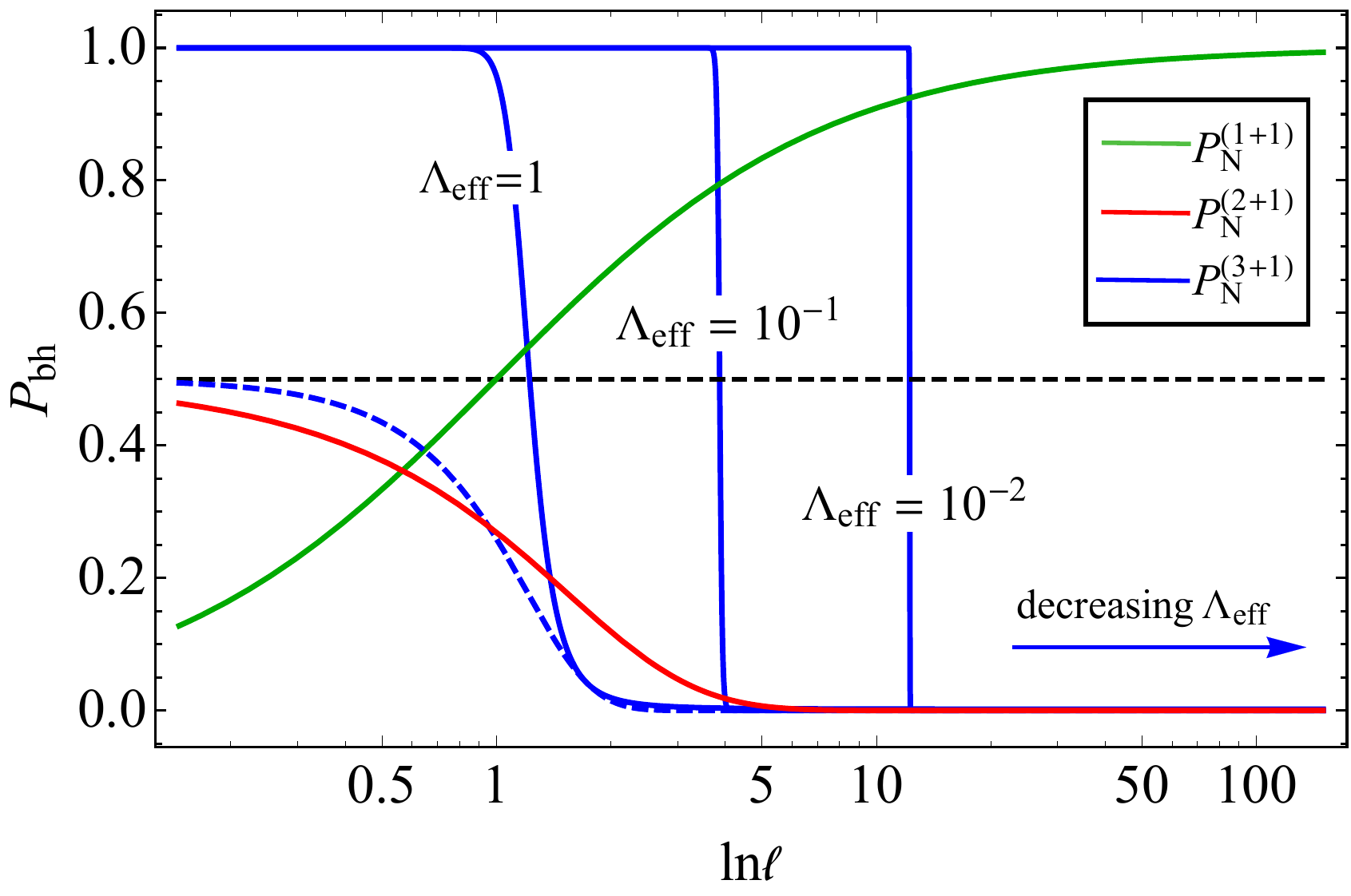}
\caption{The black hole probabilities  \textit{vs} the logarithm of the de Sitter radius $\ell$ are plotted for $|\Lambda_2| = \Lambda _3 = \Lambda$ and for $G_{2}=G_{3}=G_{\mathrm{N}}=\mu_0=1\,$. The blue dashed line stands for the conventional Nariai black hole probability.} 
\label{fig:Fig3}
\end{figure}

More importantly, the above calculation offers a possible history of a lower dimensional Universe in relation to the spontaneous production of PBHs, as seen in Fig~\ref{fig:Fig2} and~\ref{fig:Fig3}, where for ease of notation we introduced the de Sitter radius $\ell$, to express the cosmological constants $\Lambda_2$, $\Lambda_3$ and $\Lambda$ as $1/\ell^2$.  

Specifically, when the Universe was in its effective ($1+1$) dimensional era, PBHs would have been plentifully produced since the rate \eqref{rate1} is unsuppressed. Conversely, during the ($2+1$) dimensional phase, the  rate \eqref{rate2}
is exponentially suppressed but to a lesser extent than the conventional Bousso-Hawking result. For $G_3^2 \Lambda_3 \sim 1 $ the black hole nucleation is not negligible.
Hence, a lower dimensional Universe could enhance the population of PBHs relative to standard ($3+1$) scenario, provided that the magnitude of the cosmological constants  started with Planckian values during the Planck era and then decreased as the Universe expanded. Nevertheless, all the black holes formed before inflation would have been exponentially diluted leaving no observational traces.  The only possibility for significant effects, is that  the lower dimensional phases left an imprint of its pair production  at short scales after inflation. Being the oxidizing dimension an effective quantity related to the local fractality of the spacetime, such an occurrence might be acceptable  also in the case the topological dimension of the (ambient) spacetime is four. As a result, the scenario is compatible with the standard paradigm of the inflation.

Regarding the ($3+1$) phase, one finds the most promising results since the presence of the parameter $\Lambda_{\rm eff}$ allows to circumvent the issue of the production prior/after the inflation. The dilaton corrects the Bousso-Hawking rate with an unsuppressed part. 
The new Nariai rate leads to an unsuppressed production even at the final stages of inflation, as long as $\Lambda > 2\Lambda_{\rm eff}\,$. We must stress here that we expect for $\Lambda_{\rm eff}$ to have much lower value than those displayed in Fig.~\ref{fig:Fig2}, in order for the production to continue beyond the inflationary era. Inflation should have lasted at least for 60 e-foldings in order to solve the various cosmological problems (horizon, flatness and monopole problem), providing a lower post-inflationary value for the de Sitter radius of the order $\ell_{\rm end} \gtrsim e^{60} L_{\text{P}}\,$. In the plot we just give some arbitrary values for $\Lambda_{\rm eff}$ to see the new corrected behavior of the ($3+1$) rate, even post-inflationary unsuppressed rates are clearly admissible.
Conversely if the value of $\Lambda$ enters the domain $\Lambda_{\rm eff} < \Lambda < 2 \Lambda_{\rm eff}$ before the Universe reaches the late time expansion, the rate becomes suppressed at late times with no observational consequences. 

On the ground of this reasoning, one can find a concrete result stemming from our investigation. By using the current value of the cosmological constant and by invoking the Universe stability, i.e., no black hole production, the present-day value of $\Lambda_{\rm eff}$ can be constrained by using the relation $\Lambda_{\rm eff} < \Lambda < 2 \Lambda_{\rm eff}$. Being  $ \Lambda\simeq 2.888\times 10^{-122}$ in Planck units \cite{Planck15}, one finds: 
\begin{equation}
1.444\times 10^{-122}<\Lambda_{\rm eff}<2.888\times 10^{-122}.
\end{equation}

\section{Final Remarks}
\label{sec:conclusion}

In this paper, we assumed that the Universe underwent a dimensional oxidation before it reached its current form. Specifically, at very high temperatures  the Universe had ($1+1$) dimensions, later on as it cooled down, it became ($2+1$) dimensional and finally, it took its known ($3+1$) dimensional form. Taking this for granted, for each of the above phases, we calculated the probability of Schwarzschild-de Sitter black holes to be nucleated inside an expanding de Sitter background. We did this by using the no boundary proposal to calculate the associated gravitational instantons. 

In the ($1+1$) phase the existence of a dilaton is necessary. In ($2+1$) we saw that, in order for black holes to exist, the dilaton should be trivial and higher order derivative terms should appear in the action. In its current ($3+1$) phase, there is no necessity for the dilaton to exist, however if it does it solves many of the shortcomings that general relativity possesses. That is why we considered a specific case of the Horndeski action that contains a Galileon field which is symmetric under the shift transformation $\psi \rightarrow \psi + \text{constant}$. In cosmology this model is very successful in describing the late-time acceleration of the Universe with a self-accelerating solution given by the scalar field. 

We calculated the Nariai pair creation rate in each phase. Pre-inflationary production is not of interest because pre-inflationary nucleated black holes would have been washed away by inflation. If there is, however, an imprint of the lower dimensional phase after inflation, then the production of lower dimensional Planckian relics mayb still continue until today. There is an additional production from the ($3+1$) phase, depending on the values of the coupling parameters and specifically for $\Lambda > 2 \Lambda _{\rm eff}\,$. Then the rate is unsuppressed for this range of values even at the final stages of inflation. However, the decrease of $\Lambda$ should start satisfying the relation $\Lambda_{\rm eff} < \Lambda < 2 \Lambda_{\rm eff}$ as we approach present times.

Interestingly the proposed investigation offers a concrete result. The condition $\Lambda_{\rm eff} < \Lambda < 2 \Lambda_{\rm eff}$ can be applied to the current Universe to obtain compelling constraints for the John term of the Horndenski action.

\acknowledgments
K.F.D would like to thank the Institute of Space Sciences and Astronomy of the University of Malta where part of this work was conducted.
The work of P.N. has been partially supported by GNFM, the Italian National Group for Mathematical Physics.
The work of A.G.T. has been supported by the GRADE Completion Scholarships, which are funded by the STIBET program of the German Academic Exchange Service (DAAD) and the Stiftung zur F\"orderung der internationalen wissenschaftlichen Beziehungen der Johann Wolfgang Goethe-Universit\"at.   
The authors acknowledge the networking support by the COST Actions GWverse CA16104 and CANTATA CA15117.

\end{document}